\documentclass[12pt]{article}
\usepackage{graphicx}
\begin{document}

\centerline{Gamete recognition and complementary haplotypes in sexual
Penna ageing model}

\bigskip

\centerline{S. Cebrat and D. Stauffer$^*$}

\bigskip
\noindent
Department of Genomics, 
Wroc{\l}aw University, ul. Przybyszewskiego 63/77, 51-148 Wroc{\l}aw, Poland

\medskip
\noindent
$^*$ Visiting from Institute for Theoretical Physics, Cologne University, 
D-50923 K\"oln, Euroland

\bigskip
{\small
In simulations of sexual reproduction with diploid individuals, we introduce 
that female haploid gametes recognize one specific allele of the genomes as a
marker of the male 
haploid gametes. They fuse to zygotes preferrably with male gametes having
a different marker than their own. This gamete recognition enhances the 
advantage of complementary bit-strings in the simulated diploid individuals, at
low recombination rates. Thus with rare recombinations the bit-string evolve
to be complementary; with recombination rate above about 0.1 instead they 
evolve under Darwinian purification selection, with few bits mutated.}
\bigskip


\bigskip

Darwinian evolution is usually thought to come from a counterplay of 
two opposing trends: Survival of the fittest purifies the genome from
bad mutations; and random accidents produce new bad mutations. The 
Penna ageing model of mutation accumulation, as reviewed in detail
elsewhere \cite{books}, is one such example. Bad mutations are stored
in strings of bits, they reduce the survival chances of the adults, 
and new bad mutations may happen at birth. Then a realistic mortality
is obtained, increasing exponentially with adult age, for both asexual
and sexual reproduction.

\begin{figure}[hbt]
\begin{center}
\includegraphics[angle=-90,scale=0.5]{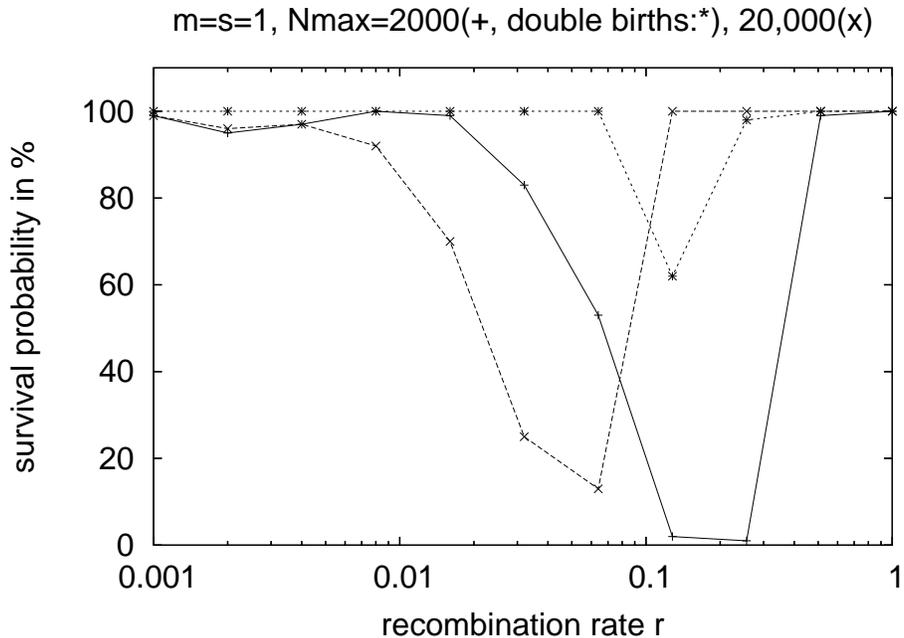}
\end{center}
\caption{Number,from hundred runs, of samples where the population survived
for at least $t = 10^5$ iterations. The + refers to our standard parameters,
for + the populations were ten times higher, and for the stars instead the 
birth rate was doubled from 2 to 4. 
}
\end{figure}

\begin{figure}[hbt]
\begin{center}
\includegraphics[angle=-90,scale=0.5]{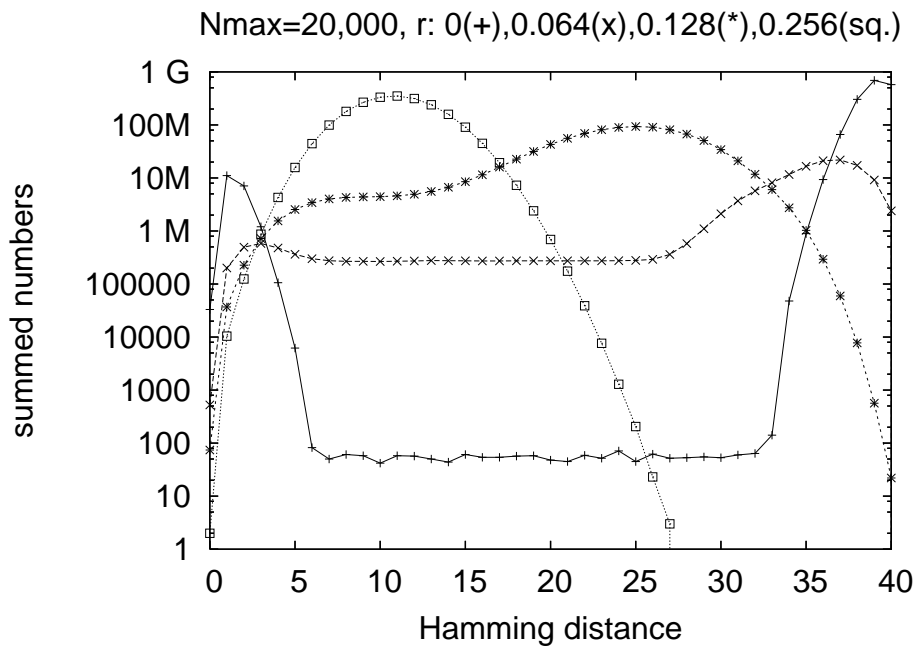}
\end{center}
\caption{Distribution of Hamming distances for one of the simulations 
of the previous figure.
}
\end{figure}

\begin{figure}[!hbt]
\begin{center}
\includegraphics[angle=-90,scale=0.4]{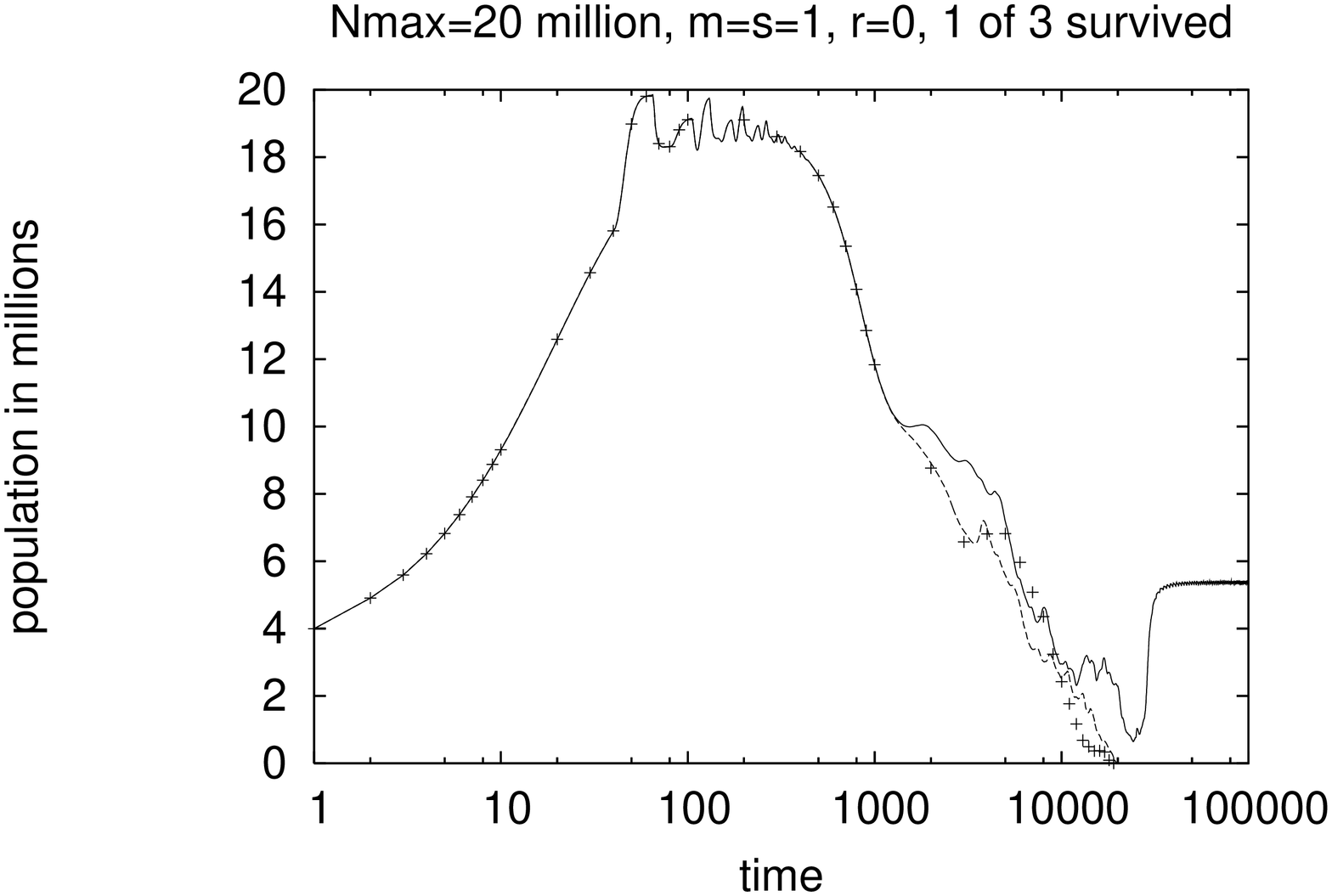}
\includegraphics[angle=-90,scale=0.4]{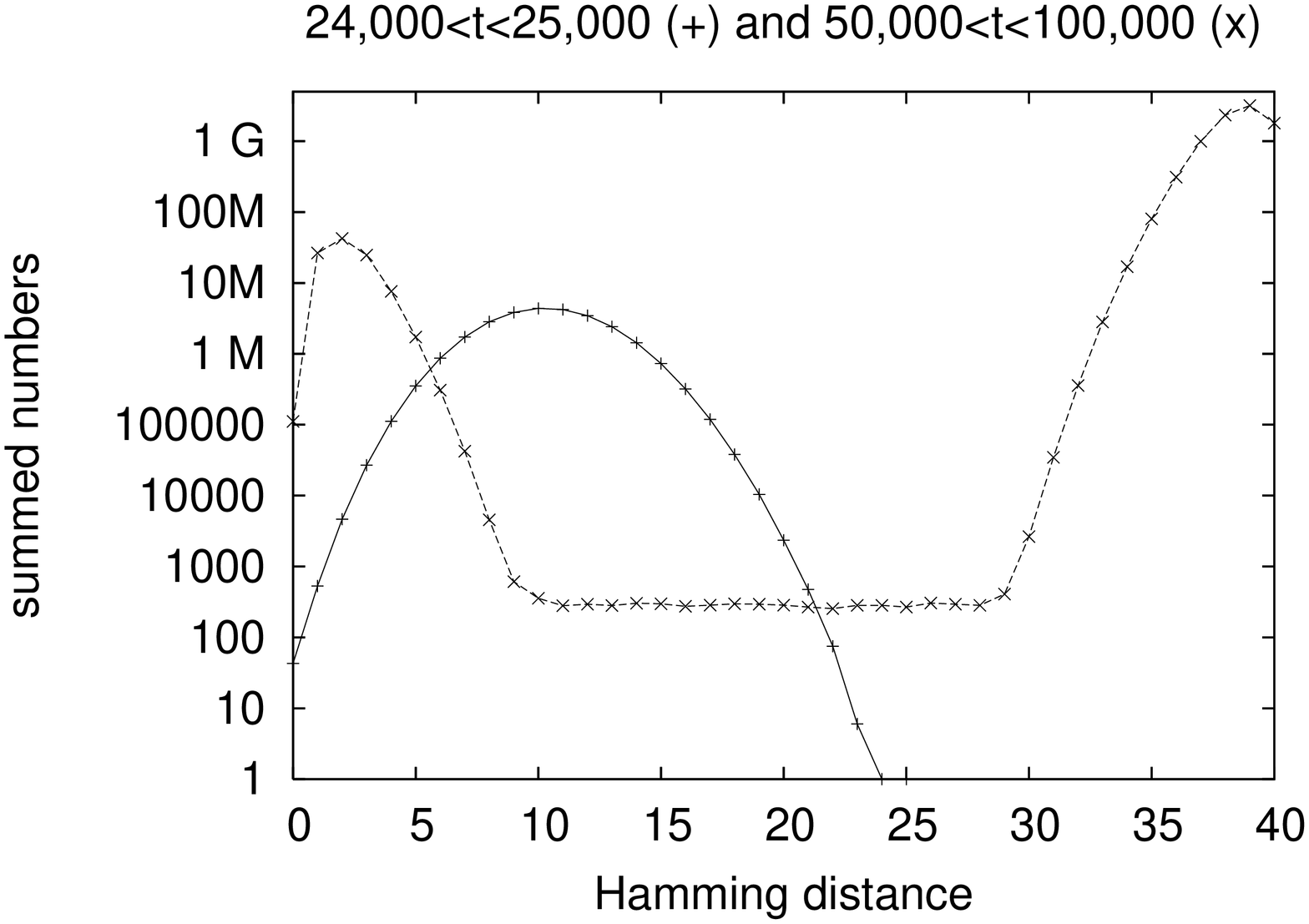}
\end{center}
\caption{Time variation of the population (part a) and of the Hamming distance
of a large population.
}
\end{figure}

\begin{figure}[hbt]
\begin{center}
\includegraphics[angle=-90,scale=0.5]{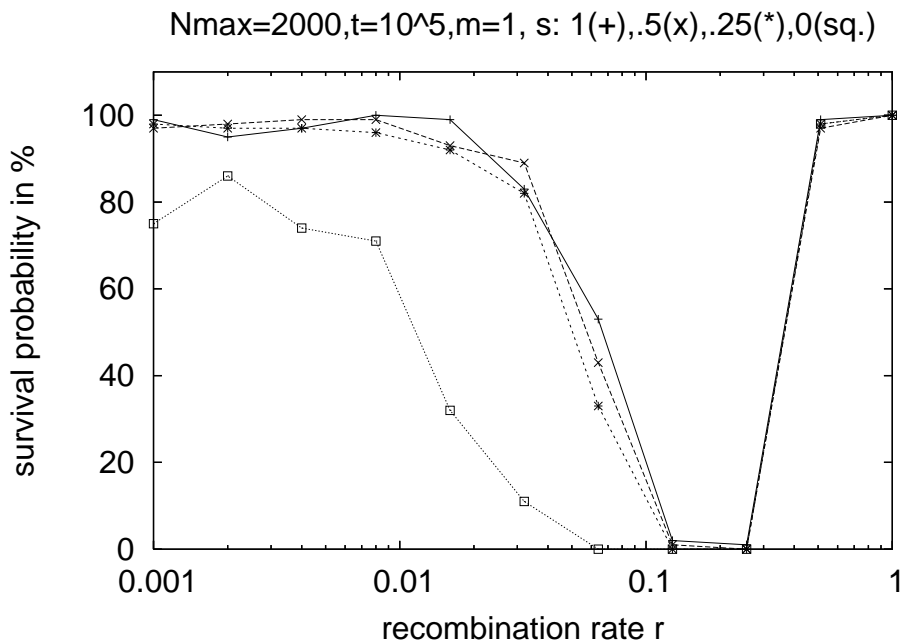}
\end{center}
\caption{Dependence of the gamete recognition probability $s$.
}
\end{figure}

\begin{figure}[hbt]
\begin{center}
\includegraphics[angle=-90,scale=0.5]{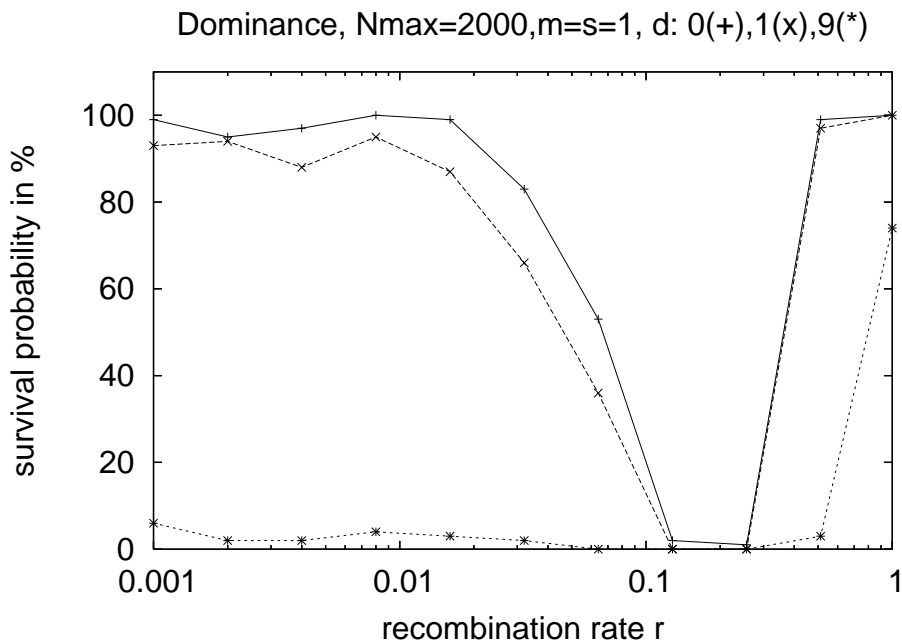}
\end{center}
\caption{Dependence on the number $d$ of dominant bits.
}
\end{figure}

\begin{figure}[hbt]
\begin{center}
\includegraphics[angle=-90,scale=0.5]{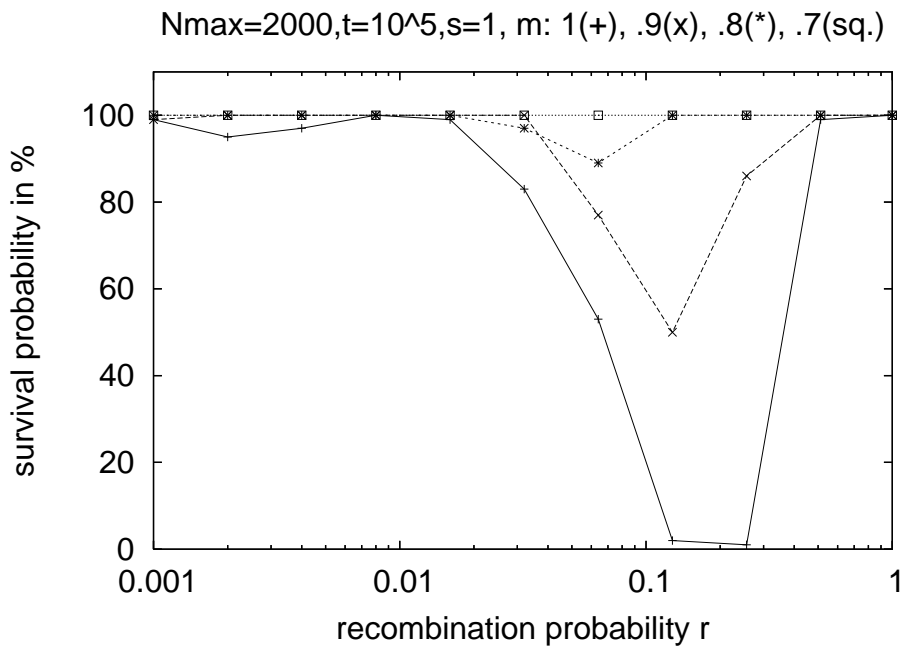}
\end{center}
\caption{Dependence on the mutation probability $m$ per iteration.
}
\end{figure}

\begin{figure}[hbt]
\begin{center}
\includegraphics[angle=-90,scale=0.5]{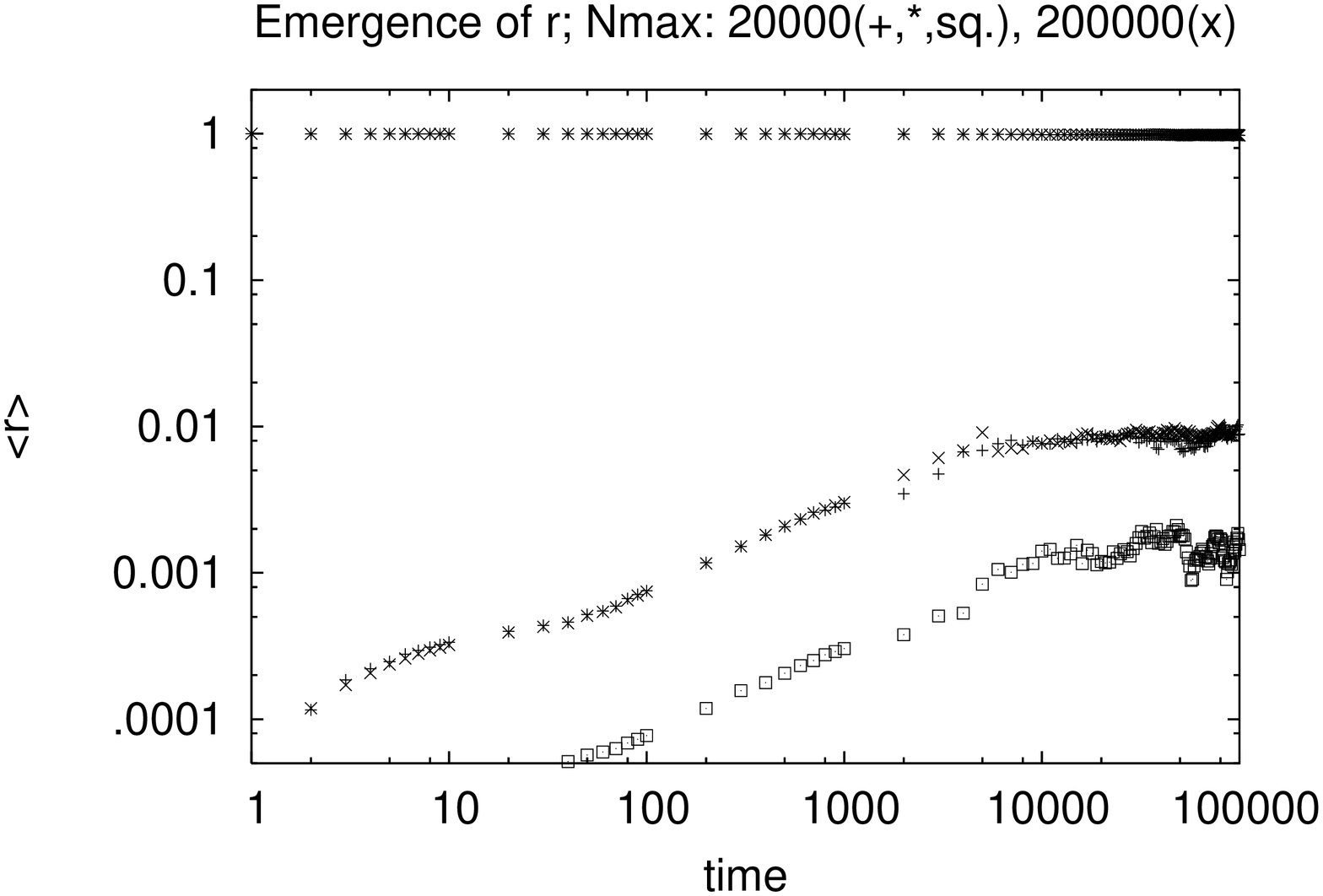}
\end{center}
\caption{Emergemce of a self-organising average recombination rate.
Results for $N_{max}=2$ million were similar (not shown).
}
\end{figure}

Recently, however, several simulations of sexual reproduction
\cite{zawierta,bonkowska,pekalski,pmco} found another strategy: the 
two bit-strings of sexual (``diploid'') animals may become complementary.
Wherever one bit-string has a zero, the other has a one, and vice versa.  
If all mutations are recessive, and no genetic loci are dominant, then 
this combination of bit-strings is as favourable as if both bit-strings
would contain only zeroes. Most of the bit-strings in the equilibrium 
population then are then divided into two groups A and B, such that each
diploid individual has one A and one B bit-string. After sexual reproduction
mixes paternal and maternal genome, only those zygotes with one A and one B 
bit-string can survive long. This ideal picture is made fuzzy due to 
new mutations (at birth), some dominant loci, and recombination (crossover)
of the two bit-strings. Thus the emergence of complementary bit-strings
from initially mutation-free genomes happens at low but not at high
recombination rates $r$ \cite{zawierta,bonkowska,pekalski,pmco,waga}. 
At high $r$ above about 0.1, instead purification dominates, than means 
most bits are zero.

In the present work we continue Ref. \cite{zawierta} and introduce selection of 
good male gametes in sexual reproduction. It was claimed \cite{pekalski,jacob}
that recognition of the Major Histocompatibility Complex (MHC) guides females
in the selection of males. This, however, seems to be unrealistic; we 
know that the type of car, the size of the residence, and the bank accounts
are the crucial selection criteria and were ignored in all simulations. Instead
we deal with recognition and selection on the level of gametes having only one 
bit-string (``haploid''). We assume that the haploid female gamete (``ovum'') 
recognises a ``marker'', i.e. a part of the genome in the haploid male gamete 
(``sperm''), before fusion into a zygote takes place. Thus a B ovum prefers
to select an A sperm, and vise versa, somewhat similar to antibody-antigen 
reactions in immunology. This marker corresponds to the ``driver'' in 
\cite{zawierta}.

More precisely, the first of the $L$ bits in a bit-string determines gamete
preference. Before fusion, with probability $s$ a female gamete tries to fuse
with a male gamete having the opposite first bit. If the ovum has zero as first
bit, it tries to find a male gamete having one as the first bit, 
and vise versa. If the first attempt is unsuccessful, it is repeated
until a gamete with opposite first bit is found. All repetitions involve
the male gametes of the same male partner, but these gametes undergo the
usual mutations before they are tested.  

In our standard case, the carrying capacity $N_{\max}$ for the Verhulst 
factor (applied to births only, not to adults) was 2000; the bit-strength
had a length $L = 64$; at each iteration $128/L$ births happened per adult
female; the minimum reproduction age was $5L/8$; $T = 1$ active mutations
killed the individual, at birth one random irreversible mutation (from 
good = 0 to bad = 1) was tried in each gamete; if the selected bit was already
one, nothing changed. Survival was tested for
$t = 10^5$ iterations (in one iterations all individuals aged by one time unit 
and could reproduce, if old enough). Gamete recognition and selection is 
switched on after one quarter of the simulation time $t$.  For the standard 
aspects of the Penna model we refer to \cite{books} and numerous articles in 
this journal, like \cite{zawierta}. We now look at the dependence on the crossover
(recombination) probability $r$ between 0 and 1, and systematically vary 
various input parameters. We check how many, of 100, samples survive up
to the end of the simulation, this number is then the survival probability
in percent. 

To measure the degree of complementarity we looked at the Hamming distance
which is the number of positions on which the bits of the male and the female
gamete differ in a position-by-position comparison at the moment of fusion
into a zygote. Only the $5L/8$ (usually 40) bits for young age were compared
since the bits beyond the minimum age of reproduction are mostly mutated to one
after equilibrium has been reached. Thus the Hamming distance varies from 0
(full agreement) to 40 (full complementarity). We sum it up over the second 
half of the simulation, i.e. over 50,000 iterations. 

Fig.1 shows our standard case (plus signs) as well as one with higher population
and another one with higher birth rate. For the higher population, Fig.2 gives 
the resulting distribution of Hamming distance. As in \cite{zawierta}, 
at intermediate $r$ survival is difficult. For smaller $r$, complementarity
dominates with Hamming distances near 40 (with about half the bits set to zero =
correct allele), while for higher $r$ the two bit-string are much more similar 
and most bits are zero. Higher populations had difficulties to 
survive long enough but after two failed attempts we succeeded in one with 
about 10 million individuals, Fig.3. The Hamming distances are small
and the population low shortly before gamete recognition is switched on
while the population rapidly increase when gamete recognition is switched on.

When we reduce the gamete recognition probability $s$, survival at low $r$ 
becomes more difficult, Fig.4; the corresponding curves are more equidistant on 
the left part if instead of the number of the surviving samples we plot the
average population (not shown). Increasing the number $d$ of dominant loci from 
the standard case $d=0$ has little effect for $d=1$ but is disastrous for 
$d=9$ at low $r$, Fig.5. Surprisingly, a slight reduction of the mutation
probability $m$ from the standard $m=1$ increases drastically the survival 
at intermediate $r$, Fig.6. 

The following results are not shown as figures:
When we reduce the length of the bit-strings (from 64 to 32 and 16), or
the minimum reproduction age (from 40 to 24), the minimum of the survival 
probability becomes less pronounced and shift to smaller $r$. 
When the observation time is increased from $10^5$ to $10^6$, the survival
chances go down drastically except for $r=1$; in principle, finite populations
(with $10^3$ individuals like ours) always vanish if simulated over sufficiently
long times. The distinction between small and large $r$ in the distribution of 
Hamming was always observed. 
If the lethal threshold $T$ for the number of mutations is increased to 2 and 3,
from 1, all samples survived, and the complementarity was less pronounced.

Finally we looked at the emergence of an optimal recombination rate $r$ in the
standard way \cite{ito}. Each individual has its own $r$, initially 0 for all,
or 1 for all. Then at birth sons inherit their $r$ from the father and daughters
from the mother, apart from a random change by $\pm 0.001$. Figure 7 shows 
the average (over all females) first to increase from zero up to about 0.01
(middle curve); if instead we start from $r=1$ for all, the average remains
above 0.95: upper data. (If we reduce the change from $\pm 0.001$ to $\pm 
0.0001$, also the final average is reduced by this factor ten: lower curve.) 
Thus, the recombination rate $r$ ``wants'' to stay either at zero or at one, 
but mutations hinder it to stay at these ideal values. 

In summary, we confirmed the separation of two evolutionary strategies 
\cite{zawierta}, bit-string complementarity at low recombination rate
and purification selection (few mutations) at high recombination rate, 
for different parameters. In particular, Fig.4, our newly
introduced gamete selection enhances the usefulness of this complementarity.

We thank COST P10 for supporting the visit of DS.

\end{document}